\documentclass[twocolumn,amsmath,amssymb,prl]{revtex4}

\usepackage{graphicx} 
\usepackage{dcolumn}  
\usepackage{bm}       

\begin{document}

\title{Quasiparticle bands in cuprates by quantum chemical methods:\\ 
towards an \textit{ab initio} description of strong electron correlations} 

\author{L. Hozoi and M. Laad}
\affiliation{Max-Planck-Institut f\"{u}r Physik komplexer Systeme,
             N\"{o}thnitzer Str.~38, 01187 Dresden, Germany}

\date{\today}

\begin{abstract}
Realistic electronic-structure calculations for correlated
Mott insulators are notoriously hard.
Here we present an {\it ab initio} {\it multi}configuration scheme that adequately 
describes strong correlation effects involving Cu $3d$ and O $2p$ electrons in 
layered cuprates. 
In particular, the O $2p$ states giving rise to the Zhang-Rice band are explicitly
considered.
Renormalization effects due to nonlocal spin interactions are also treated 
consistently. 
We show that the dispersion of the lowest band observed in photoemission is reproduced
with quantitative accuracy. 
Additionally, the evolution of the Fermi surface with doping follows directly
from our {\it ab initio} data.
Our results thus open a new avenue for the first-principles investigation of the
electronic structure of correlated Mott insulators.
\end{abstract}


\maketitle

Angle-resolved photoemission spectroscopy (ARPES) 
has proved to be an invaluable tool for
probing the correlated electronic structure and the one-particle spectral 
function, $A({\bf k},\omega)$, of interacting many-body systems such
as the $3d$ transition-metal (TM) oxides.
In addition to the ubiquitous strong $d$-$d$ Hubbard correlations, for 
late TM oxide compounds like the cuprates and nickelates, the
electron-removal spectra are strongly affected by O\,$2p$\,--\,TM\,$3d$ charge
transfer and rehybridization effects.
Detailed analysis of the photoemission data, see Ref.~\cite{SA_84},
led to the conclusion that the valence-band states at low binding energies
actually have predominant O $2p$ character in nickelates and cuprates.
Of particular interest is the nature of the first electron-removal states in
the high-$T_c$ superconductors.
For the (Mott) insulating parent compounds, the formal valence states of copper
and oxygen are Cu$^{2+}$ $3d^9$ and O$^{2-}$ $2p^6$, respectively, with one
$3d$ hole at each Cu site.
In a seminal work, Zhang and Rice (ZR) \cite{ZR_88} argued that extra holes 
doped into the CuO$_2$ layers mainly populate the ligand $2p$ levels and
give rise to impurity-like states involving singlet coupling with Cu $3d$ holes.
This was indeed confirmed by both quantum chemical 
wave-function based calculations on finite clusters \cite{CuO_martin_93,CuO_calzado_00_01}
and periodic density-functional calculations within the LDA+$U$ approximation
\cite{CuO_FMp_OKA}.
Matrix elements associated with the hopping of the $2p$ hole to neighboring
plaquettes were also computed in Ref.~\cite{CuO_calzado_00_01}.
However, the moderate size of the clusters
(up to four CuO$_4$ plaquettes \cite{CuO_calzado_00_01}) precluded a study of the effect of the  
antiferromagnetic (AF) spin background on the effective hoppings.

Calculations based on $t$\,-$J$ model Hamiltonians predict that the dispersion of 
the quasiparticle bands is strongly renormalized by the underlying AF interactions
to $O(J)\!\ll\!t$ \cite{book_fulde95}.
A relatively small width for the bands close to the Fermi level was indeed observed
by ARPES investigations.
ARPES thus plays a central role in studying the dressing of charge carriers by
interactions with the spin (and lattice) degrees of freedom
\cite{CuO_damascelli_rev,CuO_fink_rev}.
In addition, these measurements provide detailed information on the doping
dependence of the Fermi surface (FS).
In particular, the ``nodal-antinodal'' dichotomy in underdoped cuprates
\cite{CuO_tanaka06_kanigel06} was also emphasized by very recent Raman scattering
data \cite{CuO_LeTacon_06}: 
the intensity along the nodal direction scales with $T_{c}$, while the antinodal
signal actually anti-correlates with $T_{c}(x)$.
The details of the FS are important for the proper interpretation of other 
experiments as well, such as the inelastic neutron scattering \cite{CuO_norman_07},
and put strong constraints on theoretical models for cuprates.

To date, the above issues have been systematically addressed by cluster
dynamical mean-field investigations on {\it effective} one-band 
Hubbard models \cite{CuO_civelli_07,CuO_maier_05,CuO_kotliarHPs_06}.
A first-principles correlated electronic-structure calculation, however, is
still quite a distant goal, especially for late TM oxides.
Here, we present an {\it ab initio} quantum chemical study, describing how the 
detailed quasiparticle dispersion can be obtained with quantitative accuracy.
To this end, multiconfiguration (MC) calculations were performed on clusters
that are large enough to account for spin interactions in the neighborhood of
the ZR hole. 
Renormalization effects on the nearest-neighbor (NN) effective hopping were previously
investigated at the {\it ab initio} level in Ref.~\cite{CuO_hozoi_07}.
In this Letter, we show that accurate estimates for the longer-range hoppings
are crucial for explaining both, the dispersion observed by ARPES measurements,
as well as the Fermi surface, in detail.

The \textit{ab initio} calculations are performed on finite fragments cut
from the periodic system.
In constructing the MC wave-function, the orbitals of each finite cluster 
are partitioned into three different sets:
the inactive levels, doubly occupied in all configurations,
the virtual orbitals, empty in all configurations, and the active orbital
set, where no occupation restrictions are imposed.
This type of MC wave-function is also referred to as a complete-active-space
(CAS) wave-function \cite{QC_book_00}.
In the case of an undoped cluster, for example, with formal Cu $3d^9$ and
O $2p^6$ occupations of the valence levels, the most natural
choice for the active space is to include all copper $3d_{x^2-y^2}$
orbitals, akin to the wave-function in the one-band Hubbard
model at half filling.
Still, all integrals, including the on-site and longer range Coulomb interactions,
are here computed {\it ab initio}.
Although electrons in lower (inactive) energy levels, e.g., O $2s$ and $2p$,
do not participate explicitly in the construction of the many-body
expansion of the MC wave-function, they are still allowed to readjust to
fluctuations within the active orbital space.
If extra holes are created, the active space must be enlarged with orbitals
from the inactive orbital group.
A single doped hole requires one orbital to be transferred from the inactive
to the active space.
Since different electronic states have in many cases very differently
shaped orbitals, it is important to express each state in terms of its own,
separately optimized set of orbitals.
For the lowest electron-removal state, for example, the orbital added to
the active space turns into a ZR $p$-$d$ composite \cite{ZR_88} in the 
variational calculation, localized on a given CuO$_4$ plaquette.
The character of the active orbitals may change substantially, however, when
separate CAS self-consistent-field (CASSCF) optimizations are performed
for higher-lying hole states, see below.

\begin{figure}[b]
\includegraphics*[angle=270,width=0.90\columnwidth]{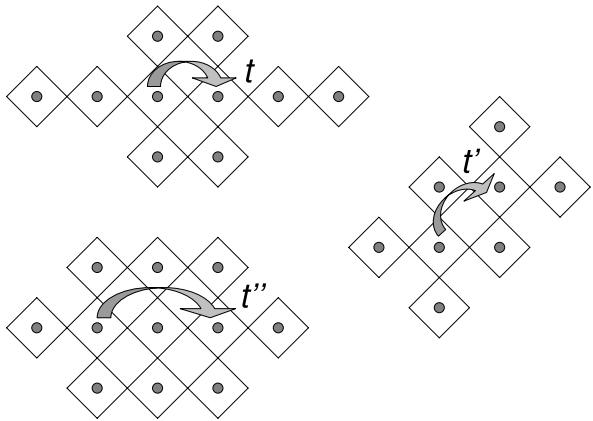}
\caption{Sketch of the finite clusters employed for the calculation of
the effective hoppings of the ZR-like hole.}
\end{figure}

The effective hoppings (and quasiparticle bands) are computed by using the overlap,
$S_{ij}$, and Hamiltonian, $H_{ij}$, matrix elements between $(N\!-\!1)$
wave-functions having the extra hole located on different plaquettes ($i$,$j$,...)
of a given cluster.
Each of these $(N\!-\!1)$ wave-functions, $|\Psi_{i}^{N-1}\rangle$, is obtained
by separate CASSCF optimizations.
A similar approach was previously applied to simpler, noncontroversial systems
such as MgO (see, e.g., \cite{MgO_hozoi_07} and references therein).
For degenerate ($H_{ii}\!=\!H_{jj}$) hole states:
$t = 1/2(\epsilon _j - \epsilon _i) = (H_{ij} - S_{ij} H_{ii}) / (1 - S_{ij}^2)$,
where $\epsilon _i$ and $\epsilon _j$ are the eigenvalues of the $2\!\times\!2$ 
secular problem.
For non-degenerate ($H_{ii}\!\neq\!H_{jj}$) hole states,
$t = 1/2[(\epsilon_j - \epsilon_i)^2 - (H_{jj} - H_{ii})^2]^{1/2}$.
The overlap and Hamiltonian matrix elements between the individually optimized,
nonorthogonal CASSCF wave-functions are obtained by State-Interaction (SI)
\cite{SI_Malmqvist_86} calculations.

The short-range magnetic correlations are described 
by including in our finite clusters extra CuO$_4$ units around those plaquettes
involved in the hopping process.
We call those latter plaquettes ``active''. All-electron basis sets of triple-zeta
quality  \cite{note_BSs} are applied for the Cu and O ions of the active
plaquettes. The core electrons of the remaining ions
in each cluster are modeled by effective core potentials
(ECP's) \cite{ECPs_CuO}. That this approximation works very well was
shown for the NN hoppings in Ref.~\cite{CuO_hozoi_07}. To describe
the finite charge distribution at the sites in the
immediate neighborhood of the cluster, we model those
ions by effective ionic potentials, see \cite{TIPs_CuSr}. Beyond
these neighbors, we use large arrays of point charges that
reproduce the Madelung field within the cluster region.
The various clusters employed in our calculations are
sketched in Fig.~1. Apical ligands are explicitly included
in our calculations only for the active plaquettes. Other
apex oxygens are represented by formal point charges.
All calculations were performed with the {\sc molcas} package \cite{molcas6}.
The structural data reported for the La$_{1.85}$Sr$_{0.15}$CuO$_4$ compound
\cite{LaCuO_xrd_cava87} were used, with an in-plane lattice constant
$a\!=\!3.78$\ \AA.

\begin{table}[t]
\caption{Occupation numbers (ON's) of the Cu $d_{x^2-y^2}$, Cu
$d_{3z^2-r^2}$, in-plane $\sigma$-type O $p_x/p_y$, and apical (ap.) O $p_z$
atomic orbitals for the ZR and the $d_{3z^2-r^2}$ hole states.
For undoped plaquettes, the ON's of the $\sigma$ $2p$ orbitals are
$\approx\!1.85$.
The ``missing'' charge is in the Cu $4s$ and $4p$ orbitals.}
\begin{ruledtabular}
\begin{tabular}{lcccc}
ON's                &$d_{x^2-y^2}$ &$d_{3z^2-r^2}$ &$\sigma$ $p_x/p_y$ &ap.~$p_z$ \\
\colrule
ZR hole             &1.05          &2.00           &1.60               &1.95  \\
$d_{3z^2-r^2}$ hole &1.40          &1.15           &1.70               &1.85  \\
\end{tabular}
\end{ruledtabular}
\end{table}

The dispersion of $d$-like states on a square lattice is given by the
following relation:
$\epsilon(\mathbf{k}) = -2t\,(\cos k_xa  + \cos k_ya) - 4t'\cos k_xa \cos k_ya
                        -2t''(\cos 2k_xa + \cos 2k_ya)...$\,,
where $t$, $t'$, $t''$ are the hopping integrals between NN, second-NN, and third-NN 
sites and each effective site is a whole CuO$_4$ plaquette.
As discussed in Ref.~\cite{CuO_hozoi_07}, a hole in an O $2p$ 
$\sigma$-type orbital induces ferromagnetic (FM) correlations among the 
NN Cu $3d$ spins.
These spin polarization ``tails'' around each $2p$ hole might actually play 
a role in pairing.
When moving through the AF background, the hole must drag along this spin
polarization cloud at nearby Cu sites, which gives rise to a substantial
reduction of the effective hopping parameters.
The spin polarization and relaxation effects at nearby sites are treated 
consistently in our approach because a separate optimization is performed for 
each particular $|\Psi_{i}^{N-1}\rangle$ configuration.
A renormalization by a factor of four was found for the NN effective hopping,
from $t_{0}\!=\!0.5\!-\!0.6$ eV to $t\!=\!0.135$ eV \cite{CuO_hozoi_07}.
The {\it renormalized} values for $t'$ and $t''$, obtained by CASSCF and SI
calculations on clusters like those shown in Fig.~1, are $-0.015$ and $0.073$
eV, respectively.
Surprisingly, $t''$ turns out to be about half of the renormalized
NN hopping $t$.   
The effect of such a large third-NN hopping is illustrated in the upper panel
of Fig.~2.
It leads to a weakly dispersing band at the ($\pi$,0) point, with a small 
``dip'' around that region, and a local energy maximum not far from the
($\pi/2$,$\pi/2$) point.
For comparison, we also plotted in Fig.~2 the dispersion along the nodal
direction for $t''\!=\!0$ (the dotted curve).
 
An important issue that must be clarified is whether there is significant
mixing between the lowest-energy hole state, the ZR-like state, and
hole states at higher binding energies.
LDA+$U$ calculations \cite{CuO_FMp_OKA}
indicate that the separation between the ZR configuration and the next
electron-removal state is of the order of 0.1 eV.
This second state is related to the creation of a $3d_{3z^2-r^2}$ hole.
Using model Hamiltonian calculations, Eskes and Sawatzky \cite{CuO_z2_eskes91}
predicted a separation of about 1.5 eV.
They also pointed out that the largest mixing should occur along the
antinodal (0,0)--($\pi$,0) direction.

\begin{figure}[b]
\includegraphics*[angle=0,width=0.95\columnwidth]{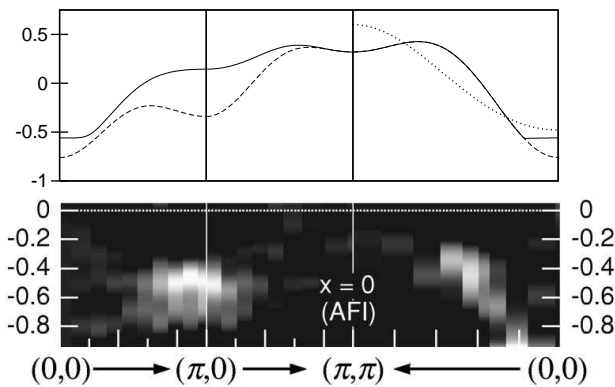}
\caption{(upper panel:) The lowest electron-removal band, without including
the interaction with the $d_{3z^2-r^2}$ hole state (dashed line) and
after including this interaction (solid line).
There is no mixing along the nodal direction.
Close to the $\Gamma$ point, the two bands cross.
For simplicity, only the lowest electron-removal band is shown.
The reference is the value of the on-site Hamiltonian matrix element
of the ZR state, see text.
The dotted line along the nodal direction corresponds to $t''=0$.
(lower panel:) ARPES data obtained for La$_{2}$CuO$_4$ by Ino {\it et al.}
\cite{CuO_ARPES_ino00}.
Units of eV are used in both plots.}
\end{figure}

Quantum chemical calculations \cite{CuO_martin_93} were previously
used for the interpretation of the optical absorption \cite{CuO_dd_perkins98}.
The shoulder at $1.5\!-\!2.0$ eV was assigned to transitions between the
$d_{x^2-y^2}$ and $d_{3z^2-r^2}$ levels \cite{CuO_dd_perkins98}.
Regarding the electron-removal spectrum, our CASSCF calculations
indicate a separation $\Delta\epsilon\!=\!0.60$ eV between the ZR and the 
$d_{3z^2-r^2}$ $(N\!-\!1)$ states.
Occupation numbers of the relevant atomic orbitals (Mulliken charges, see 
\cite{QC_book_00}) are listed in Table I.
For the $d_{3z^2-r^2}$ hole state, the two holes on a given plaquette are
high-spin coupled. 
Due to the mutual Coulomb repulsion and strong $p$-$d$ hybridization, the 
$(x^2\!-\!y^2)$ hole has large weight at the ligand sites.
The NN $d$-$d$ spin correlations are AF for this higher excited state.
Highly accurate results for the on-site energies of the ZR and $d_{3z^2-r^2}$
hole states require calculations beyond the CASSCF level.
It is known, however, that the corrections brought by more sophisticated
techniques such as the multi-reference configuration-interaction method
\cite{QC_book_00} to the on-site CASSCF {\it relative} energies are usually not 
larger than $20\%$ in TM oxides. 
Illustrative in this respect are the results of Martin and Hay
\cite{CuO_martin_93} for the $d$-$d$ excitations in undoped
clusters.
For the inter-site matrix elements, the corrections to the CASSCF results
are much smaller, see, e.g., the discussion in Refs.~\cite{CuO_hozoi_07,MgO_hozoi_07}.

\begin{figure}[t]
\includegraphics*[width=0.95\columnwidth]{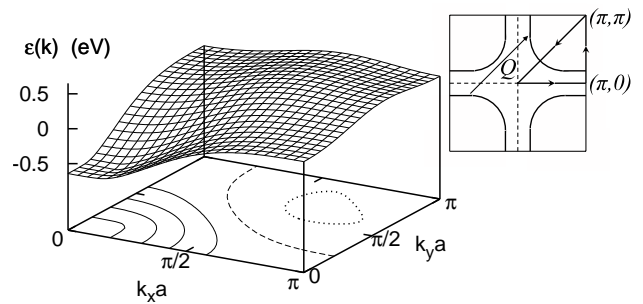}
\caption{Three-dimensional plot of the ZR-like quasiparticle dispersion,
including the effect of the ZR--$z^2$ interaction.
Constant-energy contours are also drawn in the figure.
The dashed line resembles the ``FS'' measured
in doped La$_{2}$CuO$_4$ \cite{CuO_ARPES_yoshida03} and other cuprates
\cite{CuO_damascelli_rev}, see the sketch in the inset.}
\end{figure}

We found that the mixing between ZR and $z^2$ hole states on the same plaquette
is negligible.
However, the {\it inter-site} matrix elements are large. 
The SI calculations indicate a NN effective hopping of 0.15 eV between the ZR
and $z^2$ $(N\!-\!1)$ states. 
We denote this quantity as $t_m$.
The $\mathbf{k}$-dependent term describing the mixing between the two bands,
ZR and $z^2$, now reads $t_m (\cos k_x a - \cos k_y a)$.
The NN hopping between degenerate $z^2$ hole states is much smaller,
$t_{z^2}\!=\!0.01$ eV.
It is now trivial to diagonalize the $\mathbf{k}$-dependent $2\!\times\!2$ matrix
to yield the renormalized quasiparticle bands.
The resulting dispersion of the ``($x^{2}\!-\!y^{2}$)'' band is compared to the
ARPES data in Fig.~2.
It is known that as the $\Gamma$ point is approached, spectral weight is
transferred from the lowest binding-energy peak to a higher-energy, rapidly
dispersing feature whose origin is a matter of active debate.
This higher-energy feature, not visible in the lower panel of Fig.~2
but clearly observed in Ref.~\cite{CuO_WF_ronning05} and other studies, 
and the spectral-weight transfer at the $\Gamma$ point are not addressed in this
work.
Remarkably, in all other respects, our {\it ab initio} data agree very well with 
the dispersion of the lowest ARPES band.
In particular, the flat dispersion around ($\pi$,0), the maximum near $(\pi/2,\pi/2)$,
the asymmetry along the nodal direction with respect to the ($\pi/2$,$\pi/2$) point
\cite{CuO_ARPES_ino00,CuO_ARPES_yoshida03}, and a renormalized bandwidth 
of nearly 1 eV are all faithfully reproduced in the theoretical results.
Further, in the lightly doped ($x\!\ll\!1$) cuprates, the added holes would 
immediately populate states near ${{k}}_{\mathrm{n}}\!=\!\frac{1}{a}(\pi/2,\pi/2)$:
this would give rise to nodal quasiparticles whose weight grows linearly with
$x$.
In a rigid-band picture, the quasiparticle dispersion plotted in Fig.~3 suggests
the formation of small hole pockets in the deeply underdoped regime (see the dotted
contour), as inferred from recent magnetoresistance oscillation measurements
\cite{CuO_HP_07}, the appearance of a holelike cylindrical ``Fermi surface'' at
intermediate dopings, and a change to an electronlike FS in the overdoped samples
(the full lines in Fig.~3).
Interestingly, a nearly rigid shift of the chemical potential was clearly observed
with doping in several cuprates \cite{CuO_ChemPotShift_arpes,CuO_ChemPotShift_XPS}.
The renormalized FS which would result at intermediate dopings (the dashed
curve in Fig.~3) shows all the characteristics of that measured experimentally
\cite{CuO_damascelli_rev,CuO_ARPES_yoshida03}:
correlations induce pronounced flattening of the dispersion near the antinodal
points, while electronic states in the nodal region remain dispersive.
That the ZR--$z^{2}$ mixing is important is shown by the fact that it drastically
reduces the dispersive character
close to $k_{\mathrm{an}}\!=\!(\pi/a,0)$,
bringing the computed ZR dispersion as well as the FS in much closer agreement
with the experiment.
The band at higher binding energy, associated with the charge transfer from 
the $d_{3z^2-r^2}$ level to the in-plane $p$-$d$ orbitals, was not
resolved experimentally.
More conclusive elaboration of this aspect requires one to explicitly treat the 
dynamical spectral-weight transfer between the ZR and higher excited states: 
whether this can lead to an understanding of ``waterfalls'' in the dispersion 
at higher binding energy is out of scope of the present work.

Our theoretical findings have other important implications.
Very recent electronic Raman scattering measurements \cite{CuO_LeTacon_06} clearly
show the ``nodal-antinodal'' dichotomy in cuprates.  For small $x$, the 
nodal lineshape exhibits a low-energy increase, roughly $\sim\!x$, while
the antinodal lineshape is almost independent of $x$, fully in agreement with 
our results above.  
Further, the shape of the derived FS implies existence of the ``nesting''
vectors, ${\bf Q}$, joining the flat-band parts of the {\it renormalized}
dispersion, see the inset in Fig.~3: these are important when a description
of the inelastic neutron scattering (INS) measurements is attempted
\cite{CuO_norman_07}.
ARPES, Raman, and INS results can thus be reconciled with our calculated FS.

To conclude, quantum chemical MC calculations were employed
to study the lowest electron-removal states in an undoped layered cuprate.
The method is sufficiently general and readily adaptable for studying the
correlated band structure of other $3d$-oxides of intense current interest.
The ZR physics and farther spin interactions around a ZR hole are 
treated on equal footing in our approach, which allows an accurate description 
of the renormalization effects on the dispersion of the ZR-like quasiparticle
band.
FM correlations involving spins at nearest Cu sites and AF couplings 
beyond these nearest neighbors are both treated consistently. 
The mixing with ``triplet'' $d_{3z^2-r^2}$ hole states at higher binding energy
is also investigated.
We show that {\it all} these effects are important for describing the dispersion
of the lowest $(N\!-\!1)$ band. 
In contrast, the FM $d$-$d$ correlations induced by the ZR $2p$ hole
and the ZR--$z^2$ mixing are missing in standard $t$\,-$J$ or single-band Hubbard
models.
The agreement between our theoretical results and the available ARPES data
is remarkable.
We are able to reproduce all the important details of the experimental spectrum.  
Extension of our {\it ab intio} approach to tackle dynamical responses is an 
outstanding open problem, which we leave for future consideration.

We thank S. Nishimoto and P. Fulde for insightful discussions.


\begin{thebibliography}{9}
\bibitem{note_BSs} These are atomic natural orbital (ANO) basis sets
from the \textsc{molcas} \cite{molcas6} library, with the following
contractions
\cite{QC_book_00}: Cu ($21s15p10d$)/[$5s4p3d$] and O ($14s9p$)/[$4s3p$].
\end{thebibliography}
\end{document}